\documentstyle[twoside,fleqn,espcrc2,epsfig]{article}

\title{Semileptonic Decays of Heavy Mesons with the Fat Clover Action
\thanks{Presented by C.~DeTar.}}
\author{ C.~Bernard
\address{Department of Physics, Washington University, St.~Louis, MO 63130, USA},
T.A.~DeGrand
\address{Physics Department, University of Colorado, Boulder, CO 80309, USA},
C.E.~DeTar
\address{Physics Department, University of Utah, Salt Lake City, UT
  84112, USA},
Steven~Gottlieb
\address{Department of Physics, Indiana University, Bloomington, IN 47405, USA},
U.M.~Heller
\address{SCRI, Florida State University, Tallahassee, FL 32306-4130, USA},
J.E.~Hetrick
\address{University of the Pacific, Stockton, CA 95211, USA},
Craig McNeile
\address{Department of Math Sciences, University of Liverpool, L69 7ZE, UK},
K.~Orginos
\address{Department of Physics, University of Arizona, Tucson, AZ 85721, USA}, 
R.L.~Sugar
\address{Department of Physics, University of California, Santa Barbara, CA 93106, USA},
and D.~Toussaint$\,\null^{\rm h}$,
} 
\begin{document}

\begin{abstract}
We are studying a variety of semileptonic decays of heavy-light mesons
in an effort to improve the determination of the heavy-quark
Standard-Model CKM matrix elements.  Our fermion action is a novel,
improved ``fat'' clover action that promises to reduce problems with
exceptional configurations.  Dynamical sea quarks are included in a
mixed approach, {\it i.e.} we use staggered sea quarks and fat-clover
valence quarks.  Here we report preliminary results.
\end{abstract}

\maketitle 
\section{OBJECTIVES} We are studying the semileptonic
decays $B \rightarrow \pi \ell \nu$, $B \rightarrow D \ell \nu$, $B
\rightarrow \rho \ell \nu$, $B \rightarrow D^* \ell \nu$, and $B
\rightarrow K^* \gamma$ and the corresponding decays with a strange
spectator quark. For a companion study of purely leptonic decays, see
\cite{Gottlieb_lat99}.  The CKM matrix element $V_{ub}$, for example,
is obtained from the differential semileptonic decay rate for $B
\rightarrow \pi \ell \nu$ at total leptonic four-momentum $q$
\cite{Richman_Burchat}:
\begin{displaymath}
\frac{d\Gamma}{dq^2} = 
	\frac{G_F^2 {p^\prime}^3} {24 \pi^3} 
        |V_{ub}|^2 |f^+(q^2)|^2.
\end{displaymath}
The unknown hadronic form factor $f^+(q^2)$ is to be determined in
lattice gauge theory from the matrix element of the weak vector
current $V_\mu$,
\begin{eqnarray*}
 && \langle\pi(k)|V_\mu|B(p)\rangle = \\
 && \ \ \ \ \ \ \left(p_\mu+k_\mu-q_\mu\frac{m_B^2 - m_\pi^2}{q^2}\right) f^+(q^2) \\ 
 &&  \ \ \ \ \ \ + q_\mu\frac{m_B^2 - m_\pi^2}{q^2}f^0(q^2).
\end{eqnarray*}
\section{FAT CLOVER ACTION} Since the heavy-light meson decays involve
light quarks, it is important to choose an ${\cal O}(a^2)$ lattice
fermion implementation with good chiral properties.  To this end we
have been experimenting with an action proposed by DeGrand,
Hasenfratz, and Kov\'acs \cite{Fat_Clover_Action}, which introduces,
in effect, a cutoff-dependent form factor at the quark-gluon vertex to
suppress lattice artifacts at the level of the cutoff.  The action is
the usual clover action but with a gauge background constructed by
replacing the usual gauge links by APE-smoothed links \cite{APE_block}
with coefficient $1-c$ for the forward link and $c/6$ for the sum of
staples.  The smoothed link is projected back to SU(3).  This
smoothing process is repeated $N$ times.  For the present experiment
we take $c = 0.45$ and $N = 10$.  These values are to be kept constant
in the continuum limit, thus giving the local continuum fermion
action.  This ``fattening'' process reduces problems with
``exceptional'' configurations that obstruct extrapolations to light
quark mass \cite{Fat_Scaling,BD_fat_pert_renorm}.

\begin{figure}
 \vspace*{-25mm}
 \epsfig{bbllx=100,bblly=230,bburx=530,bbury=740,clip=,
         file=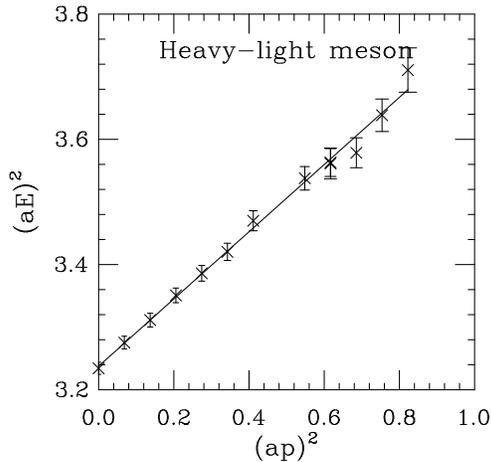,width=70mm}
 \vspace*{-5mm}
\caption{Dispersion relation for a heavy-light meson with approximate
quark masses $1.1 m_b$ and $0.5 m_s$.  The slope is $c^2 = 0.537 \pm
0.018$.
\label{fig:disp_rel}
}
\vspace*{-8mm}
\end{figure}

\section{PARAMETERS IN THE STUDY}

Calculations were done on an archive of 200 $24^3 \times 64$ gauge
configurations, generated with two flavors of dynamical staggered
quarks of mass $am_q = 0.01$ at the one-plaquette coupling $6/g^2 =
5.6$, corresponding to a lattice spacing (from the rho mass) of about
0.11 fm.  The fat clover propagator was generated for three ``light''
(spectator and recoiling) quarks and five ``heavy'' (decaying and
recoiling) quarks over a mass range $0.5 m_s < m < 1.1 m_b$.  The
coefficient of the clover term $c_{SW}$ was set to 1. The mass of the
lightest fat clover quark was adjusted to give the same pion mass as
the staggered fermion Goldstone boson.

We use the Fermilab program through ${\cal O}(a)$ for the quark wave
function normalization, including the three-dimensional
rotation\cite{Fermilab_heavy_clover} with coefficient $d_1$.
The light meson source is
placed at $t = 0$ and the heavy-light meson at $t = 32$, with
antiperiodic boundary conditions in $t$.  We treat three values of the
heavy-light-meson momentum and 21 values of the three-momentum
transfer at the current vertex.  Computations are in progress.
Results are presented for a subset of about half of the 200
configurations including only the two lightest spectator quark masses.

\begin{figure}
 \vspace*{-25mm}
 \epsfig{bbllx=100,bblly=230,bburx=530,bbury=740,clip=,
 file=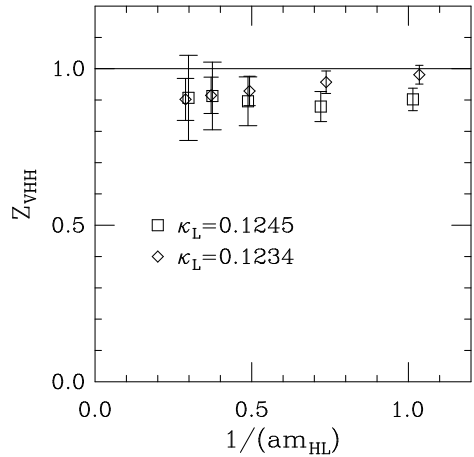,width=70mm} \vspace*{-5mm}
\caption{Diagonal vector current renormalization factor for the
heavy-light meson as a function of inverse meson mass for two choices
 of the light quark mass.
\label{fig:Z_V}
}
\vspace*{-7mm}
\end{figure}

\section{SELECTED RESULTS}

An example of the meson dispersion relation is shown in
Fig.~\ref{fig:disp_rel}.  It is quite satisfactory.

The form factor is extracted by amputating the external meson legs ---
at present, by dividing by $\exp[(E_B - E_M)t]$, where the $B$ meson
energy $E_B$ and recoil meson energy $E_M$ are taken from central
values of a fit to the corresponding two-point dispersion relations.
The diagonal vector form factor at zero three-momentum transfer gives
the vector current renormalization factor $Z_V$.  It is shown as a
function of the inverse meson mass in Fig.~\ref{fig:Z_V} for the two
currently available choices of the spectator quark mass.  We see that
this nonperturbative renormalization constant is within $10-15$\% of
unity.

\begin{figure}
 \vspace*{-25mm}
 \epsfig{bbllx=100,bblly=230,bburx=530,bbury=740,clip=,
 file=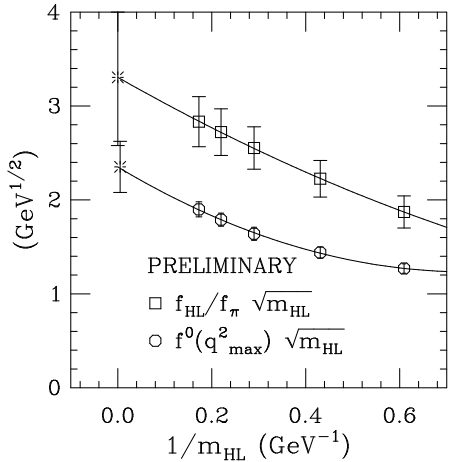,width=70mm} \vspace*{-5mm}
\caption{Test of the soft pion theorem.
\label{fig:soft_pion2}
}
\vspace*{-7mm}
\end{figure}

We test the soft pion theorem \cite{soft_pion} which states that in
the chiral limit $f^0(q^2_{\rm max}) = f_B/f_\pi$.  The same action
and configurations are used to get $f_B$ \cite{Gottlieb_lat99}.  Both
spectator and recoil quark masses ($m$ and $m^\prime$) are
extrapolated to zero.  If we use $f^0[q^2_{\rm
max}(m,m^\prime),m,m^\prime] = a + bm + cm^\prime$ we obtain
Fig.~\ref{fig:soft_pion2}, a disagreement similar to that found by
JLQCD \cite{JLQCD}.  If we include an extra term $d \sqrt{m +
m^\prime}$ as advocated by Maynard \cite{Maynard_lat98} the theorem is
satisfied, but with large extrapolated errors.  We hope our eventual
full data sample will help resolve these complexities
\cite{Lesk_lat99,Hashimoto_lat99}.

\begin{figure}
 \vspace*{-25mm}
 \epsfig{bbllx=100,bblly=230,bburx=530,bbury=740,clip=,
         file=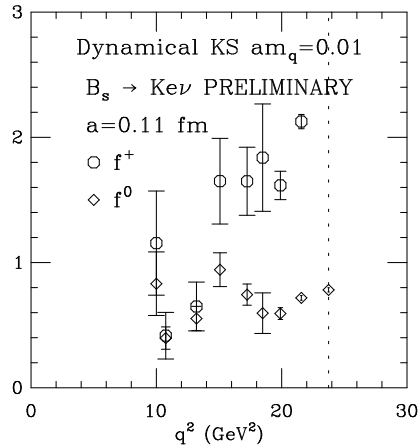,width=70mm}
 \vspace*{-5mm}
\caption{Sample form factors $f^+(q^2)$ and $f^0(q^2)$ for the process
$B_s \rightarrow K  \ell \nu$
\label{fig:Bs_to_K}
}
\vspace*{-7mm}
\end{figure}

Sample form factors for the process $B_s \rightarrow K \ell \nu$ are
shown in Fig.~\ref{fig:Bs_to_K}.
\section{DISCUSSION} Fattening has allowed us to obtain results for an
ostensibly ${\cal O}(a^2)$ action on unquenched lattices for quark
masses at least as low as $0.5 m_s$ with no noticeable trouble from
exceptional configurations.  Our experiment raises a number of
important questions: Will a one-loop-perturbative determination of
current renormalization factors be adequate?  How much fattening is
good?  Does fattening push us farther from the continuum limit for
some quantities?  Work is in progress.

This work is supported by the US National Science Foundation and
Department of Energy and used computer resources at the San Diego
Supercomputer Center (NPACI), University of Utah (CHPC), Oak Ridge
National Laboratory (CCS), and the Pittsburgh Supercomputer Center.

\begin{thebibliography}{99}
%
\bibitem{Gottlieb_lat99} Presentation by S.~Gottlieb (this conference).
%
\bibitem{Richman_Burchat} J.D.~Richman and P.R.~Burchat,
Rev.~Mod.~Phys.\ {\bf 67} (1995) 893.
%
\bibitem{Fat_Clover_Action}
T.~DeGrand, A.~Hasenfratz, and T.~Kov\'{a}cs, Nucl.\ Phys.\ {\bf B547}
(1999) 259.
%
\bibitem{APE_block}
M.~Falcioni, M.~Paciello, G.~Parisi, B.~Taglienti,
Nucl.\ Phys.\ B {\bf 251} [FS13] (1985) 624;
M.~Albanese {\it et al.},
Phys.\ Lett.\ B {\bf 192} (1987) 163.
%
\bibitem{Fat_Scaling} For a discussion of scaling and chiral zero
modes with this action, see M.~Stephenson, C.~DeTar, T.~DeGrand, and
A.~Hasenfratz, in progress (1999).
%
\bibitem{BD_fat_pert_renorm} For a discussion of perturbative
renormalization with this action, see the presentation by C.~Bernard
(this conference).
%
\bibitem{Fermilab_heavy_clover} A.~El-Khadra, A.~Kronfeld, and P.~MacKenzie,
Phys.\ Rev.\ {\bf D 55} (1997) 3933.
%
\bibitem{soft_pion} G.~Burdman and J.F.~Donoghue, 
Phys.\ Lett.\ {\bf B280} (1992) 287;
M.B.~Wise,
Phys.\ Rev.~\ {\bf D45} (1992) 2188.
%
\bibitem{JLQCD} H.~Matsufuru {\it et al.}, 
Nucl.\ Phys.\ B (Proc.\ Suppl.) {\bf 63A-C} (1998) 368.
%
\bibitem{Maynard_lat98} C.~Maynard (UKQCD), 
Nucl.\ Phys.\ B (Proc.\ Suppl.) {\bf 73} (1999) 396.
%
\bibitem{Lesk_lat99} Presentation by V.~Lesk (this conference).
%
\bibitem{Hashimoto_lat99} Plenary talk by S.~Hashimoto (this conference).
\end{thebibliography}
\end{document}